**RESEARCH**

**Open Access**

# Gamification of virtual museum curation: a case study of Chinese bronze wares

Zhaokang Li[1], Qian Zhang[2], Jiayue Xu[2], Chuntao Li[2,3*] and Xi Yang[1,3,4*]

## Abstract

Museums, which are among the most popular science institutions outside schools, are usually used to display and introduce historical culture and cultural relics to tourists. Text and audio explanations are used by traditional museums to popularize historical knowledge and science for tourists, and general interactive systems are based on desktops. This learning method is relatively boring in terms of experience. As a result, tourists have no desire or interest in actively exploring and learning about bronze ware, so they only have a basic understanding about bronze ware. Since most tourists are familiar with games, they are more likely to be attracted by game content and will actively explore and interact with it. In addition, a certain degree of reality is created by virtual reality technology and an immersive experience through head-mounted devices is provided to users. In this paper, we take Chinese bronzes as the research objects. We first use 3D laser scanners to obtain bronze models ; then, we build a virtual museum environment, and we finally design a virtual reality curation game based on this bronze digital museum. This game offers visitors an immersive museum roaming and bronze ware interactive experience. Through a combination of text, video learning, and games, visitors' curiosity and desire to explore bronze ware are stimulated, and their understanding and ability to remember bronze ware knowledge can be deepened. In terms of cultural heritage, this game is also conducive to the spread of traditional Chinese bronze culture throughout the world.

## Introduction

Bronzeware comprises utensils made of bronze alloy. It is a great invention in human history and a symbol of world civilization. It can reflect the social system, social production, cultural development, and other elements of each era in the history of each country, and it has specific and strong cultural connotations. The study of bronze culture has far-reaching significance for the development and progress of world civilization [1]. Since bronzeware is very valuable, museums usually seal and preserve it with a glass cover so that tourists cannot touch it. For some bronzeware with high historical value, museums even prohibit visitors from getting closer to observe it, and restrict the viewing position of tourists. With the emergence of digital technologies such as 3D laser scanning technology [2] and photogrammetry technology [3], cultural heritage can be digitized and then presented on a computer's 2D display by building a virtual museum through a game engine. In this way, visitors can visit the bronze artifact exhibition in a virtual museum without being restricted by time, space, or distance. However this method still has many defects: (1) it is too static and lacks the interaction design between visitors and bronze artifacts to stimulate participants' interest in exploration and desire to learn. (2) it is not immersive enough. Visitors do not have enough interest or curiosity to explore bronze artifacts in depth. The sense of connection between visitors and bronze artifacts [4] is not strong, which leads to visitors having only a superficial perception and experience with bronze artifacts. However, a series of situations can be provided by virtual reality(VR) games between users and game objects and a strong sense of immersion. Therefore, combining museum curation with VR games and designing a VR gamified bronze artifact virtual museum curation will stimulate visitors' curiosity and interest in exploring bronze artifacts and learning related knowledge because it allows visitors to immerse themselves in it.

*Correspondence:
Chuntao Li
lct33@jlu.edu.cn
Xi Yang
yangxi21@jlu.edu.cn
[1] School of Artificial Intelligence, Jilin University, Changchun 130000, Jilin, China
[2] School of Archaeology, Jilin University, Changchun 130000, Jilin, China
[3] Key Laboratory of Ancient Chinese Script, Cultural Relics and Artificial Intelligence, Jilin University, Changchun 130000, Jilin, China
[4] Engineering Research Center of Knowledge-Driven Human–Machine Intelligence, MoE, Changchun 130000, Jilin, China

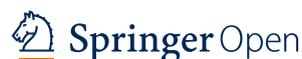

This interaction model originated from drama and is called "breaking the fourth wall"; in drama, this means breaking the invisible fourth wall between the actors on the stage and the audience (the other three walls are the stage scenes). The actors interact with the audience so that the audience becomes part of the stage play, from which the comedy and drama of the play can be enhanced. Research shows that this interaction method is highly effective [5] and is widely used in archaeology [6], education [7], psychology [8] and other fields, and VR technology is one of the best media to realize this interaction mode. In addition, tourists' curiosity and interest in exploration can be stimulated by adding game elements to educational knowledge [9]. Especially for children and young tourists who love to play games, game-based learning methods can be very effective in helping them learn new things [10]. Since the emergence of virtual reality and augmented reality technology, immersive VR/AR games have become more competitive, valuable and have better development prospects than nonimmersive desktop games and offline physical games in terms not only of experience and interest stimulation but also of money expenditure, distance, and other restrictions in the field of game-based learning [11].

Curatorial designers are allowed to try various arrangements and combinations of 3D objects in virtual scenes to display virtual museums. Designers create and disseminate virtual models of cultural relics of various countries to the public by combining archaeological knowledge with aesthetic pleasure. By visualizing relics through VR or AR technology, more realistic and attractive virtual museum exhibitions can be provided to the public. Visitors can interact in and easily explore these exhibitions, and visitors are allowed to observe objects from all angles in virtual museum exhibitions. VR/AR virtual museums can provide visitors with a strong sense of presence and immerse them in it. Because there is a positive correlation between presence and enjoyment, visitors will also enjoy museum exhibitions considerably [12]. In addition, many studies have shown that video games can motivate participants to explore for long periods of time [13–16] and gamification methods that add game elements to nongame environments can be used for educational interventions [17, 18]. In recent years, the combination of serious games and virtual museums has made the application of virtual museum curatorial gamification increasingly extensive; it can directly connect visitors to the content that the museum wants to present to meet their educational needs [19]. Moreover, gamified virtual museum curation can make interactions between visitors and artifacts more enjoyable and enhance the visitor experience [20].

In the long history of China, bronze culture has been one of the most important cultural heritages. China's Bronze Age began in 2000 BC and lasted until the Qin and Han Dynasties. Its time span was as long as 15 centuries. This long history has been passed down to this day and promotes the development of economic, political, cultural, and other industries. China's bronze culture has attracted an increasing number of international tourists to visit China for sightseeing to understand and learn the essence of traditional Chinese culture.

On this basis, in this paper, we designed a VR curatorial game for the purpose of learning knowledge about bronzes displayed in museums. First, we used a 3D laser scanner to scan bronzes from the Jilin University Museum and Changchun Museum and then completed the data collection of the 3D models of the bronzes. We then used the Unity engine to build a virtual museum scene and input the collected bronze model data and the text knowledge corresponding to each bronze. Finally, we employed an interaction design. We designed three game levels with gradually increasing difficulty. Each level consists of a roaming scene and a game scene. After visiting the bronzeware in the roaming scene of the current level, tourists can enter the game scene of this level. Through fun games, they can review the bronzeware viewed in the roaming scene and their corresponding bronzeware knowledge to deepen their understanding, that is, "learn first, play later". Only when a tourist has an acceptable correct rate and reaches the target can they pass one level and enter the next.

To verify the effectiveness of our VR-gamified bronze museum curation system, two user experiments are conducted by us. In the first experiment, we found 40 volunteers to try our system and then asked them to complete the System Usability Scale(SUS) questionnaire. Finally, we calculated the average score of all of the volunteers to evaluate the usability of the system. In the second experiment, we selected 40 volunteers and divided them into two groups. The first group of 20 volunteers experienced the 2D desktop version of the bronze museum curation, and the second group of 20 volunteers experienced the gamified VR bronze museum curation system. Each volunteer in the two groups was required to take a test after the experience, and the exam questions for everyone were the same. When all the volunteers finished answering, SPSS system is used to compare the scores of the two groups of volunteers and verify the effectiveness of our system in improving their ability to learn about bronze knowledge and remembering what they have learned.

Accordingly, the following main contributions are made in this work to better broadcast and help everyone understand China's traditional bronze culture.

(1) We scanned 200 Chinese bronzes from Jilin University Museum and Changchun Museum and generated corresponding 3D bronze models. We then built a digital museum environment to display these bronze models.
(2) We designed many interactive methods and functions that can improve the user experience, including the use of lasers to check text panels and user interfaces flexibly, the layout of teleport areas and teleport points to achieve rapid movement, and the ability to touch/grab/rotate bronze ware.
(3) By leveraging professional archaeological knowledge, we designed a VR gamified bronze museum curation that helps users better learn bronze knowledge.

## Related work on VR cultural heritage displays

Because of its high value and possible particularity of cultural heritage (geographic location, material, weight, and area), bronzeware is not easy to explore or directly touch, pick up, and observe. In reality, immersive experience cannot be provided to museums tourists. However, with the advent and continuous development and breakthrough of virtual reality technology, digitized cultural heritage combined with virtual reality can allow tourists to visit and interact with cultural heritage at their convenience without being restricted by region, time, space or distance [21, 22]. This method is widely recognized and used by relevant researchers [23]. In particular, during the COVID-19 pandemic, this kind of design is meaningful to the public who cannot leave their homes, and people's stress and anxiety of staying at home for a long time can be relieved [24]. Various brands of VR head-mounted displays (HMDs) and controller handles can be used to roam the digitized scene of cultural heritage without any restrictions [25, 26] or 360-degree view explore and view items of cultural heritage without blind spots [27]. This interactive mode allows tourists to interact with digitized cultural heritage in a virtual space in a way that is difficult to achieve in reality, and they can better perceive the existence of cultural heritage and establish a deep connection with it.

### VR-based virtual museums

There has been much research on the exhibition of VR virtual museums. For example, Guy et al. [28] developed Viking VR for a Viking culture exhibition area of a physical museum. It is a virtual reality exhibition that viewers are allowed to experience the sights and sounds of a 9th-century Viking camp. As part of the large museum exhibition, the experience was developed by an interdisciplinary team of artists, archaeologists, curators and researchers. The authors also explored issues surrounding interactive design for the long-term deployment of VR experiences in museums and discussed the challenges of VR creation workflows for interdisciplinary teams. George et al. [29] designed 10 different categories of virtual museums in a virtual environment, which can save tourists considerably money and time. See et al. [30] used VR technology to design a virtual world of the tomb of Sultan Hussein Shah of Malaysia and in this way, traditional cultural heritage is displayed to the world through VR. Loizides et al. [31] digitized the cultural heritage of Cyprus, presented it through a VR virtual museum, and found volunteers to experience it and conduct usability evaluations. These virtual museums can provide visitors with a strong sense of immersion and the connection between participants and cultural heritage is enhanced. However, this type of virtual museum is too static and lacks an interactive design between visitors and cultural heritage, so visitors do not feel very involved.

### VR gamified culture heritage learning

Serious VR-based games are a great way to learn about cultural heritage. Liu et al. [32] designed an educationally oriented VR excavation game called Relic VR that simulates the process of unearthing ancient cultural relics. They added an uncertainty design to the game by wrapping the 3D bronze model with a black cube to simulate soil on the bronze in a real tomb. Users must use the controller to control the shovel, hammer, and other tools in the game to dig out the complete bronze human face; users learn about the bronze in the process of continuous digging. Using VRGonizzi et al. [33] designed a 3D gamified interactive scene of the important ancient Egyptian ritual "The Road of the Dead", aiming to enhance the public's experience and understanding of ancient Egyptian culture. Laura et al. [34] designed and developed a VR archaeological field game system for teaching introductory archaeology courses at a midwestern university in the United States. These studies on VR gamification of cultural heritage knowledge demonstrate a series of interactive behaviors to stimulate the interest and curiosity of participants to explore and learn relevant knowledge about cultural heritage. However, in some of these studies on gamified cultural heritage learning, the amount of cultural heritage displayed and learned by visitors is not large, which result in insufficient experience time for visitors and visitors' desire to continue exploring new things cannot be satisfied. The design of interactive elements is not concise and clear enough, which causes high learning costs for visitors. These defects lead to low enjoyment for visitors. In addition, some studies require users to enter some personal information, which may lead to the leaks of private data.

The purpose of our research is to improve the public's understanding of Chinese bronze culture, help them attain Chinese bronze knowledge and spread Chinese bronze culture to all parts of the world. In previous related work, we found that there has been almost no research on combining Chinese bronze virtual museum curation with playing VR games or design a VR-gamified bronze curation that not only stimulates visitors' interest and desire to explore through the immersion brought by virtual reality and the interaction with bronzes but also helps visitors learn and remember bronze knowledge better through games with a minimum accuracy requirement. Therefore, we designed a multilevel VR game-based museum curation.

## Materials and methods

Unity is easy to port projects and supports multiple platforms. It is lightweight, intuitive, resource-rich and has low requirements for computers and VR hardware, which makes it relatively user friendly. Therefore, Unity is chosen to construct the museum virtual environment. The system consists of 3D bronze models, a virtual environment, and interaction behaviors. Figure 1 shows the framework of our system.

### 3D model collection

Typically, 3D laser scanning technology is used to digitize real-life objects. 3D laser scanning technology is based on the principle of laser ranging. It determines the distance of an object's surface by emitting a laser beam and measuring its round-trip time. The scan quickly and continuously changes the direction of the laser beam to scan the object in all directions, thereby obtaining dense point cloud data on the surface of the object [35, 36]. Photogrammetry technology uses not only photography technology to obtain images of the target object but also optical projection and geometric relationships to convert the two-dimensional image information of the object into three-dimensional space coordinates or actual dimensions through calculation [37]. 3D laser scanning can provide high-precision model data, capture the fine details of a scanned object, and generate detailed three dimensional point cloud data to accurately reflect the three-dimensional shape of the object. In addition, laser scanning is less affected by external lighting conditions and can work in different environmental situations. Although photogrammetry is inexpensive and can obtain data quickly, it has limitations in obtaining high-precision three-dimensional information, and ambient lighting conditions and the texture of the object surface have a great impact on the measurement results. Considering that our working environment is inside a museum and that we hope to obtain high-precision bronze models, we use 3D laser scanning technology in this work.

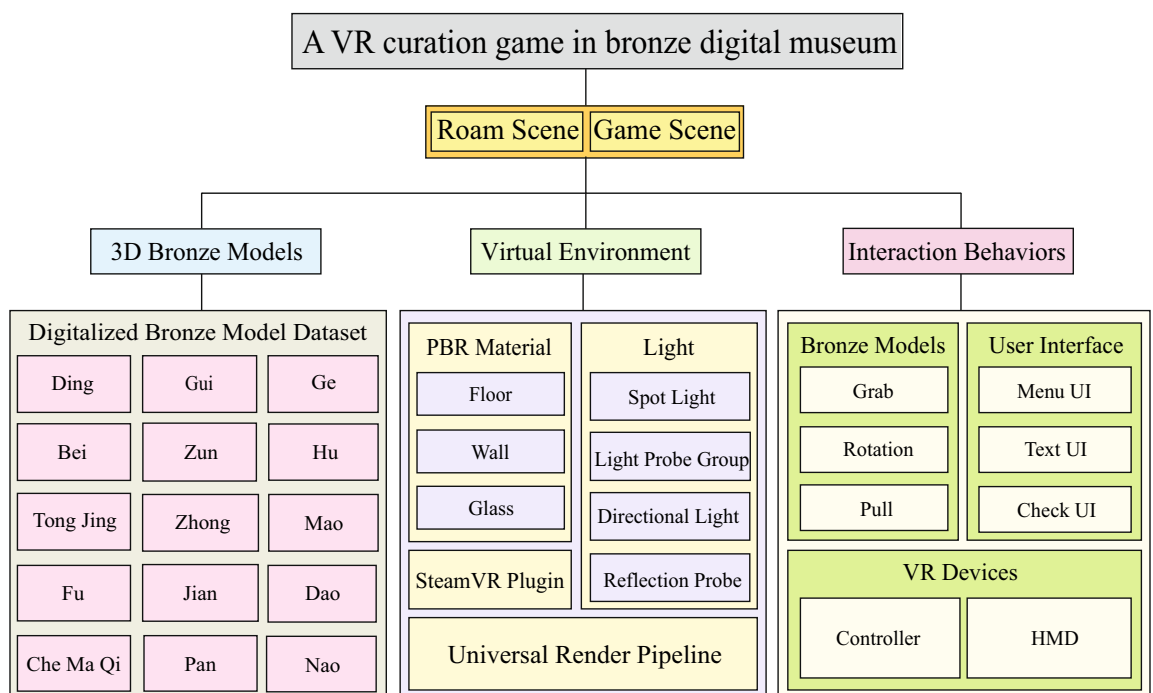

**Fig. 1** Framework of our system

The preparation of the digitized bronze models consisted of four steps: 1. obtain real bronze ware; 2. create a 3D laser scanning workflow; 3. generate and optimize a mesh model; 4. correct and simplify. First, we entered the Jilin University Museum and Changchun Museum to obtain various bronzes from various dynasties in ancient China, such as the Ding, Gui, Li, and Jue. Second, we then used 3D laser scanner (Shining 3D) to scan the bronzes and used EXScan Pro software corresponding to the scanner to generate 3D models for them. The 3D scanners we use have two modes: handheld scanning and fixed scanning. Fixed scanning can be divided into fixed turntable scanning and fixed nonturntable scanning. Handheld scanning is mainly used for artifacts that are large and heavy and, therefore, difficult to move, and its scanning accuracy is low. In contrast, the advantages of fixed scanning are a high scanning accuracy, less manual intervention, and low operating difficulty, but its application has great limitations. Fixed turntable scanning is used for small and medium-sized artifacts that are moderate in size, simple in structure, and easy to move, and fixed nonturntable scanning is suitable for artifacts that are moderate in size but complex in structure and easy to move. The handheld scanning accuracy and the density of the generated 3D point cloud are lower than those of fixed turntable scanning, but the scanning speed is higher. Our scanning of artifacts must first ensure absolute accuracy. Except for a few artifacts that were too large and too heavy, a fixed method is uesd to scan most artifacts to restore their true appearance to the greatest extent possible. During the 3D scanning process, we employed different scanning methods according to the size, weight and surface state of the artifacts to be scanned. 1. Artifact size: The largest artifact is nearly 40-cm high, the smallest is only approximately 1 cm, and the weight span ranges from dozens of kilograms to a few grams. The turntable of our equipment only bears a load of more than ten kilograms, and the length and width of the turntable surface are only a little more than 20 cm. Artifacts that are too large or too heavy could be scanned only through the handhold mode. 2. Artifact shape: Many artifacts have different degrees of concave and convex patterns on the surface, and some parts that are difficult to scan using ordinary methods, such as the hollowing out between parts and cavities in the abdomen of artifacts. The operator must constantly adjust the placement angle of the artifact so that the laser beam emitted by the laser scanner can hit the particular parts of the artifact that need to be scanned. In this process, the artifact needs to be placed in a strange posture and the operator cannot touch it. Therefore we used plasticine, wooden sticks and plastic brackets to fix the artifact so that it could be placed in the position that we needed. In addition, for bronze artifacts with complex geometric shapes or surprising carvings, we need to scan repeatedly from multiple angles to ensure that all of the details were fully captured. 3. Surface state of artifacts: The artifacts that we scanned were bronze artifacts. Most artifacts of this type have rough surfaces and complex textures, which do not require special attention. However, some artifacts, such as bronze mirrors, have smooth surfaces and metallic luster,. For this type, we needed to reduce the brightness during scanning, including blocking the light of the scanning site and lowering the brightness of the laser beam emitted by the laser scanner to restore the most realistic state of the artifact surface. For the third step, when the scanning was complete and the 3D bronze point cloud presented by the software corresponding to the 3D scanner, the software's built-in tools are used to de-noise the bronze point cloud and click global optimization. Then, we encapsulated it to generate a 3D bronze mesh model. Based on the software's built-in filling tool, we filled in any tiny missing areas in the model, clicked texture fusion, and saved the digitized 3D model data. We then used 3ds Max to make the y-axis of the bronze model's own coordinate system consistent with the normal direction of its bottom surface to add a collider to the bronze model in the digital museum. In addition, we employed MeshLab to simplify the number of meshes of the bronze model to prevent problems, such as the system delays and long response times caused by excessively large bronze models when tourists pick them up, which result in a bad experience for visitors. Figure 2(a) shows a high-polygon Ding bronzeware with high complexity. Figure 2(b) shows an optimized low-polygon Ding bronzeware. Figure 2(c) shows a high-polygon Hu bronzeware with high complexity. Figure 2(d) shows an optimized low-polygon Hu bronzeware. Figure 2(e) shows a high-polygon Zun bronzeware with high complexity.

Figure 2(f) shows an optimized low-polygon Zun bronzeware concerning the patch simplification method, we use the quadratic error metrics method [38] in MeshLab, which can quickly generate high-quality approximate

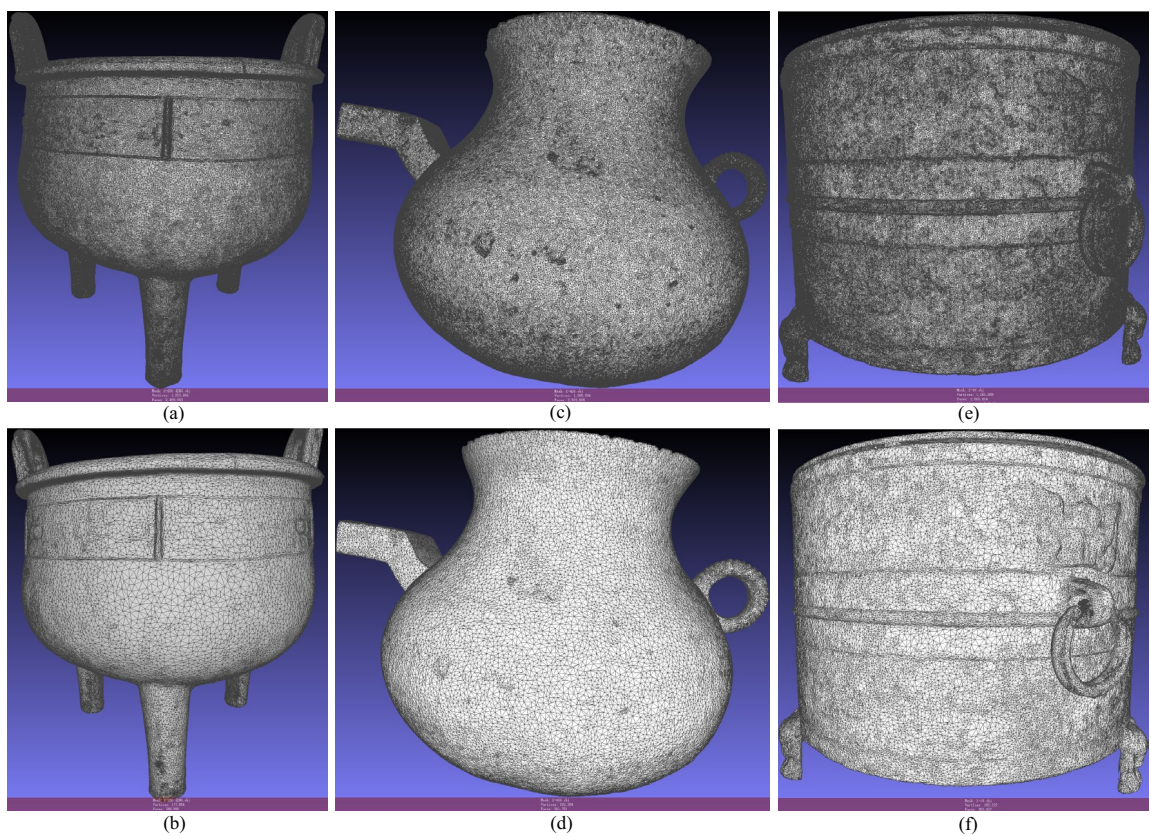

**Fig. 2** **a** Ding before optimization **b** Ding after optimization **c** Hu before optimization **d** Hu after optimization **e** Zun before optimization **f** Zun after optimization

3D models with color and texture. Although the patch simplification algorithm sacrifices some details while reducing the complexity of the model, the loss of model surface details is minimized by this method through optimizing the quadratic error metric and achieves a balance between simplification and detail preservation. At this point, we obtained 3D bronze model data that can be placed in the digital museum.

After five months of scanning, more than 200 digitized 3D models of Chinese bronzes are collected; This included the bronze data from the Archaeology and Art Museum of Jilin University, which are not available on the internet. Figure 3 shows pictures of 26 real bronzes at the Jilin University Museum. Figure 4 depicts the 26 bronze models corresponding to Fig.3 among the 200 digitized bronze models obtained through scanning.

**Constructing the virtual environment**

After obtaining a suitable empty venue model from online free resources, we used the Unity engine and chose a universal render pipeline to build the digital museum environment. It is mainly divided into three parts: material production; lighting layout; and a VR display.

*Material production*

Physically based rendering(PBR) material is a type of realistic material produced by rendering technology based on physical principles. The goal of PBR is to create realistic three-dimensional image effects to improve visual effects and the viewing experience. This material describes the optical properties of the material by using physical property parameters such as metalicity, reflectivity, and roughness. PBR materials are widely used in game development, film and television production, and other fields. Unlike texture mapping, PBR can achieve higher visual fidelity. To make the virtual museum environment more realistic and provide better visual immersion and experience, PBR materials are used for the floor and walls [39].

In this work, we used the software Substance Sampler software to complete the production of PBR materials for the floors and walls of the digital museum. This software can automatically create detailed PBR materials based on images (texture maps and photos taken) provided by users and artificial intelligence principles. Then users can determine the presentation of the final material by adjusting various parameters on the interface. After importing a picture of the floor and completing the material conversion according to artificial intelligence, we fine-tuned the parameters of different layers of the material, such as the roughness, smoothness, softness, floor shape and normal strength. Finally, we exported the finished PBR floor material to Unity. The production process of the PBR wall material was the same as above.

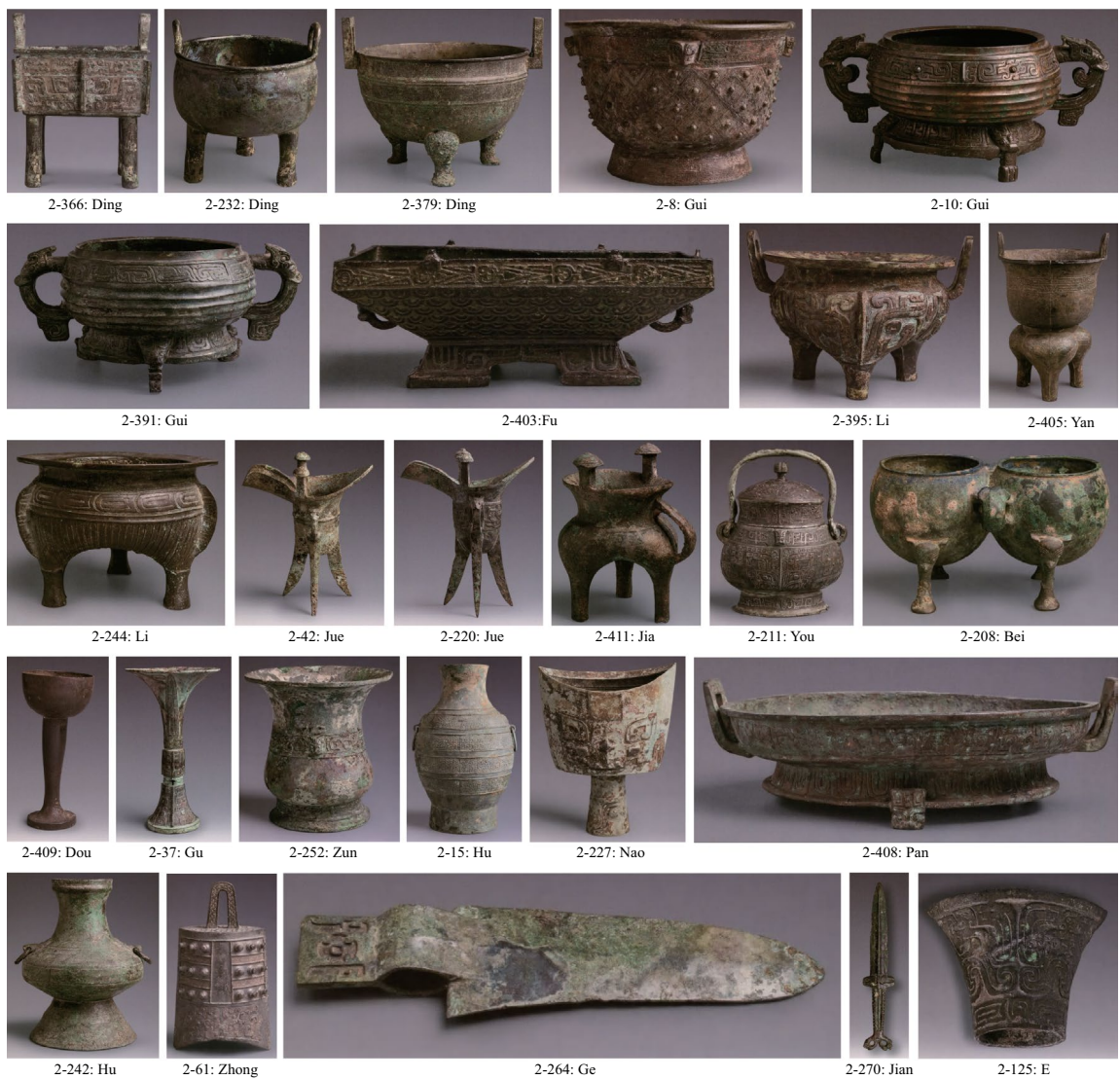

**Fig. 3** Twenty-six real bronze pictures from the Archaeology and Art Museum of Jilin University

### *Lighting layout*
The setting of lighting is the most important part of environmental construction. A good lighting layout design can make the environment more realistic and make objects more delicate and beautiful under the light. Therefore, it is necessary to make the lighting layout of the virtual bronze museum consistent with that of the real museum. We can also improve the performance of the system through light mapping technology. A light map is a map that stores the lighting information of the scene. It superimposes the lighting information on the game objects participating in the baking process in the form of a map. In this way, there is no need to perform for every frame the real-time rendering of these game objects that participate in the baking map every frame when the system is running, which results in a waste of performance [40].

Based on the lighting layout strategy of real museums [41], first, in the external environment of the digital museum the ambient light of the skybox and the parallel

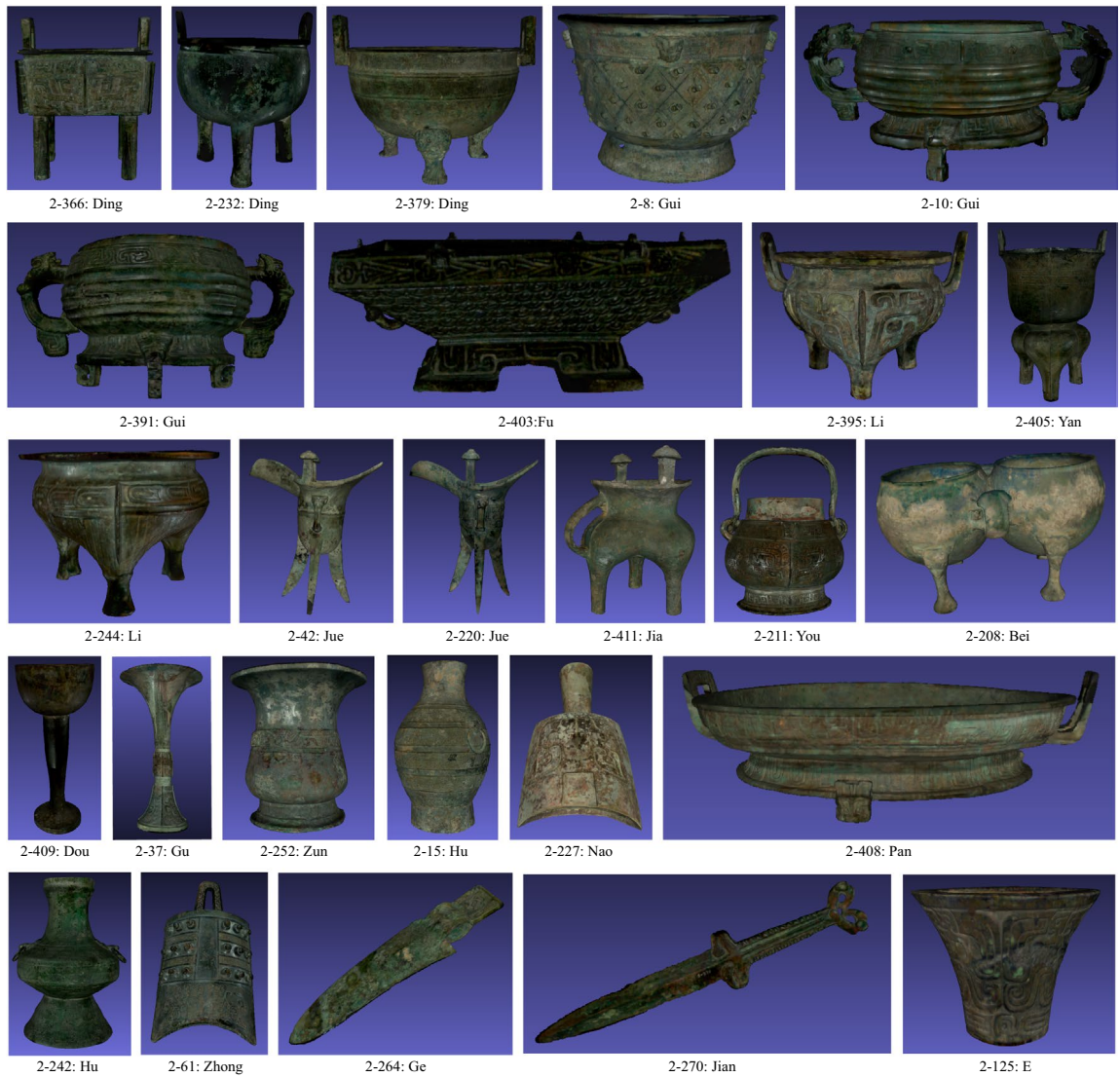

**Fig. 4** twenty-six bronze models corresponding to Fig 3

light are used as the museum's external light source to simulate real sunlight and provide light to its interior through the glass doors and the glass windows on the roof of the digital museum. Second, to make the bronze models and other exhibits in the digital museum present the lighting effects of real museum curation, we first added spotlights to directly illuminate each 3D bronze model and then added some light probes to each bronze model to obtain the lighting information of the area near each one to better match it each to its surrounding environment. Then, we added a reflection probe to the center of the digital museum to simulate the reflected light in a real museum and increase the brightness of the room. Finally, to improve performance and smoothness, we set non-interactive game objects as static objects and bake light maps on them. To simulate the light and shadow effects of people grabbing objects and moving them in the real world, the 3D bronze models were set as dynamic objects. Thus the system dynamically presents light and shadow information when a player interacts with the bronze.

*VR display*

VR is a 3D virtual environment generated by a computer. The user cannot see the real environment and is completely immersed in the virtual world presented by a head-mounted device (such as HTC Vive, Occulus Rift, Pico, etc.). High-end VR experiences can create a certain degree of reality, allowing users to enjoy realistic and immersive feelings. In recent years, an increasing number of virtual museums have chosen to combine VR technology to bring visitors a more enjoyable immersive experience [28, 42, 43]. VR devices based on the HTC Vive series use advanced spatial positioning technology, which can accurately calculate the user's relative position and rotation angle in three-dimensional space, ensuring that the user's interaction in the virtual environment is more accurate and natural, and its high-resolution and high refresh rate screen can provide users with a clearer and smoother visual experience, which enhances immersion. Therefore, in this project, HTC VIVE Pro is used as the VR hardware device.

*Game object layout*

In the three-level roaming scenes of this gamified exhibition, we displayed the bronzes on stands and then set up each stand in a circle against the wall so that visitors had a wide area to move around. In the three-level game scenes, the bronzes were placed on tables or shelves or in boxes. Since visitors will grab and observe the bronzes in each scene, to help each visitor explore and place the bronzes, and considering the diversity of visitors' heights, the tables, stands, and shelves where the bronzes are placed are set at a moderate height. In addition, we placed each text panel displaying bronze knowledge in the roaming scene on the side of the corresponding bronze and placed the text in the panel at a similar height to the bronze model so that visitors could view the bronze and text knowledge at the same time.

**Interactive design**

Interactive design is crucial to users. A novel, interesting, and easy-to-master interactive design can stimulate tourists' interest and curiosity to explore bronze ware; more importantly, it can help tourists understand and remember the bronzeware knowledge expressed. For this work we designed three game levels, with each level displaying twenty-two different bronze models. Each level consists of a roaming scene and a game scene. The interactive design of the level is divided into two parts: interactive design in the roaming scene and interactive design in the curation game scene. Common user-system interaction modes in 3D virtual environments include navigation, object selection/manipulation [44]. Therefore, in this work we established a teleportation area where visitors can freely explore, a teleportation point with text prompts, touch selection, grabbing, and rotating to view the bronze artifacts, placing bronze artifacts into containers, and other interactive behaviors that users can perform in the gamified bronze artifact curation. In addition, considering that the most effective method in terms of selection technology is to use 2D, the button layout and graphic selection design in the user interface should be minimalistic to create a user-friendly visual environment and to allow users to operate the system in an easy, simple and effective way, we designed an intuitive and easy-to-use 2D bronze artifact text knowledge viewing interface and answer submission interface for visitors [42, 44, 45]. All of the interaction behaviors are shown in Table 1.

*Interactive design of the roaming scene*

There are three roaming scenes. Except for the 3D bronze model on the rectangular stand, the other environments are exactly the same. We then designed the following interactive behaviors in the roaming scene. the roaming scene at level one is shown in Fig. 5.

I) Teleport

We used the teleportation technology provided by the SteamVR plugin to help visitors wearing a headset to move and explore the bronzes in the digital museum. Visitors can trigger the teleportation mechanism by clicking on the upper part of the touchpad of the handle controller. After selecting the target teleport point, the visitor can teleport to the target point.

**Table 1** Interaction behaviors in our system

| Interaction type | Interaction behavior | Content of this interaction |
|---|---|---|
| Interact with bronze models | Touch | Use controller to touch bronze models |
| | Grab | Use controller to grab bronze models |
| | Rotate | Use controller to rotate the bronze model after grab it |
| Interact with UI buttons | Click text panel | Use ray to click button on panel to learn bronze knowledge |
| | Click answer check panel | Use ray to click button on panel to submit answer and check accuracy |
| Interact with scenes | Teleport | Use controller to roam in scenes |

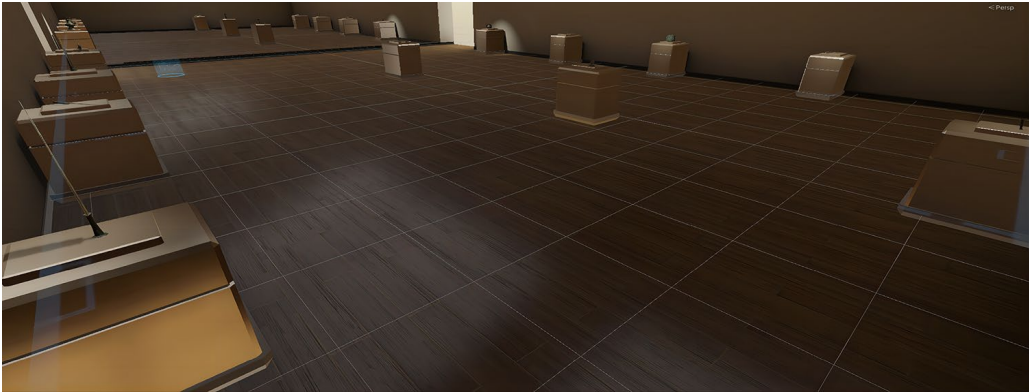

**Fig. 5** Roam scene 3

II)  Interaction with bronzeware

We added movable components to each bronze model in the scene so that visitors can grab and move each 3D bronze model when they move the hand model in the VR environment near the bronze model through the controller. When the hand model is near the bronze model, the edge of the bronze model will show a light yellow highlighted outline. Visitors can press and hold the trigger key of the handle controller to grab the bronze and then rotate and move the handle controller to view the bronze model from any angle. Additionally, they can move the bronze model in front of their eyes to explore the inscriptions, textures, and other details of the bronze more carefully. Figure 6 shows the interaction with the "Ding" bronze model.

III) Interaction button to display text

We used Unity's own UGUI system to place buttons. We placed an exclamation mark button on the side of the wall of each booth where the bronze model is placed. Visitors can click this button via light emitted from the hand model. When visitors click on the exclamation mark icon button on the wall in Fig. 7, the wall displays relevant text descriptions corresponding to the bronzes on the left side of the button, such as the age, decoration, casting technology, historical value and other bronze knowledge.

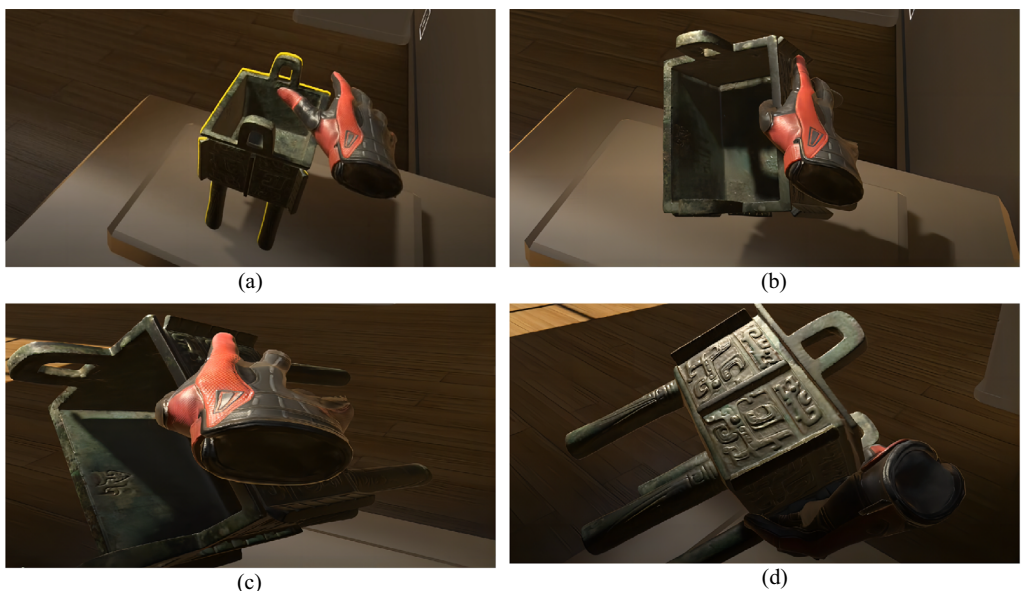

**Fig. 6** **a** By the hand near the bronze model, the edge of the bronze model has a light yellow outline. **b** Grabbing the bronze model. **c** Rotating the bronze model to view the inscription on the top of the bronze model. **d** Moving the bronze model to view the texture on the bottom clearly

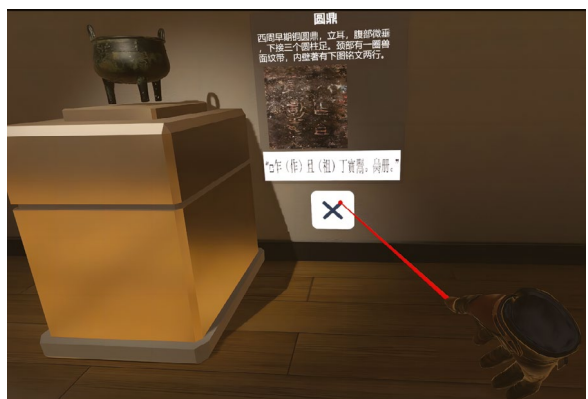

**Fig. 7** Clicking the button to view the bronze knowledge text

***Interactive design in the curation game scene***
A good game element design can stimulate visitors'™ interest and make them willing to explore and enjoy the game for a long period of time. In the design of the game content, we considered game design principles such as challenge, control, creativity, learning, and goals [46] and designed three game scenes corresponding to the three levels of roaming scenes mentioned above. After visitors pass the game of the current level, they can enter the roaming scene of the next level through the "next level" teleport point. When visitors enter the game scene from the roaming scene of the current level, the specific game process is as follows. After reading the game rules on the wall, visitors put the bronzeware into the container (shelf, table, box) that they think is correct. After the placement is complete, they can click the "Yes" button on the answer submission interface to submit the placement result. Then, the system calculates the accuracy of the placement result based on the accuracy detection mechanism. If the accuracy reaches the predesigned value of this level, the teleport point leading to the next roaming scene level will be opened, and visitors can enter the next level roaming scene. If the accuracy does not reach the predesigned value of this level, then the teleport point leading to the next level roaming scene is not opened. Visitors must modify the placement of the bronzes and try to submit the placement answers until the accuracy reaches the standard. When they cannot pass the current level for any reason, they can also return to the roaming scene of this level through the "Return to Roaming Scene" teleportation point, review the knowledge related to bronze models, and then return to the game scene of this level again to place bronzes, but the placement of bronzes before visitors return to the roaming scene will not be saved, and they must start all over again.

*ITeleport*   This is the same as the teleport part of the "Interactive design in the roaming scene" section.

 *Interaction with bronzeware*

*Game scene 1: category* The game objects in the game scene of the first level are composed of twelve bronze models selected in the roaming scene of this level, which are shelves in four categories, namely, bottles, tripods, Ge, and Gui, and a table. Twelve bronze models are placed on a table in the center of the room. Visitors need to put three bronze models on each shelf that they think belong to that category. After the placement is completed, visitors can submit the current placement of the bronzes through the hold button of the handle controller. If the accuracy rate is higher than 80%, then the roaming scene of the next level can be entered through the teleport point in the room. The game scene of the first level is shown in Fig. 8(a).

*Game scene 2: purpose*   The game objects in the game scene of the second level consist of 13 bronze models selected in the roaming scene of this level and 5 small round tables with words printed on them, namely eating, war, wine vessels, musical instruments, and sacrifices, and a long table and a chair. Twelve bronze models are placed on the long table on one side of the room, and one bronze model is placed on a chair as a display. On the other side of the room are five small round tables. Visitors need to select 10 bronze models from the 12 bronzes on the long table and place 2 bronze models on each small round table that they think are suitable for specific purposes. After the placement is complete, visitors can submit the current placement of the bronzes through the hold button of the handle controller. If the accuracy rate is higher than 90%, then the roaming scene of the next level can be entered through the transfer point in the room. The game scene of the second level is shown in Fig. 8(b).

*Game scene 3: dynasty*   The game objects in the game scene of the third level are composed of nine bronze models selected in the roaming scene of this level and three boxes with labels of Shang Zhou, Han, and Wei Jin respectively. Nine bronze models are placed on the display booths in the room. Visitors need to place the nine bronzes on the booths into the boxes that they think are correct. The rules of this level are displayed on the side of the wall. After the placement is complete, visitors can submit the current placement of the bronzes through the hold button of the handle controller. Visitors can pass this level only if the correct rate is 100%. After passing this level, the visitor passes the entire game. The game scene of the third level is shown in Fig. 8(c).

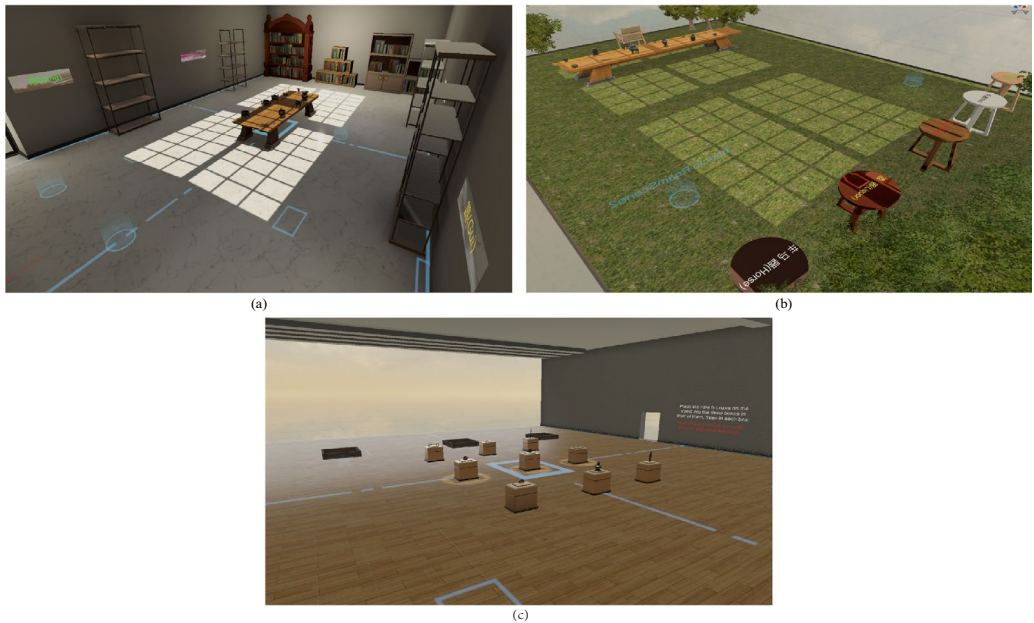

**Fig. 8** Three game scenes. (a) Game scene 1: Category. (b) Game scene 2: Purpose. (c) Game scene 3: Dynasty

*Correct rate check mechanism*    After the visitor clicks the hold button of the handle controller, the answer submission interface shown in Fig. 9a appears. The visitor clicks the "Yes" button on the interface to trigger the accuracy detection event. This accuracy detection mechanism is compared with a specific value A based on the distance between the bronze and the container of the correct category (the shelf in Game scene 1, the round table in Game scene 2, and the display cabinet in Game scene 3). The bronzes placed by tourists are detected one by one and if the distance between the currently detected bronze and the container of the correct category is less than A, then the bronze is determined to be placed in the correct container. After all of the bronzes participating in the detection in the current game scene have been detected, if the visitor's accuracy rate is higher than the predefined values for the particular level stated in the previous section, then congratulations for passing the level as shown in Fig. 9b appear, and the teleport point leading to the next level opens. If the accuracy rate does not reach the predefined standard, then the please correct answer interface in Fig. 9c appears.

**Results**

Based on the above mentioned digital 3D bronze model, virtual museum environment, and interactive design, we finally built a system consisting of three roaming scenes and three game scenes and used an HTC VIVE PRO as a VR-supported hardware device to enter the digital museum for testing.

To evaluate whether our system can achieve the expected results, 40 people are invited to experience it. The composition of the participants included 32 masters degree students and 8 doctoral students, including 20 majoring in computer science, 19 majoring in archaeology, and 1 majoring in chemistry. The experience process for each participant was as follows. First, after putting on the head mounted display and holding the two handle controllers, the participants enter the interaction example scene that comes with the SteamVR plugin. Next, we explain how to perform teleport, adjust the perspective and grab objects. When participants are proficient in the current operation, they can proceed to the next operation. After learning the three operations, the participants will enter the bronzeware digital museum and we explain how to view the text knowledge corresponding to the bronzeware and how to enter the game. Finally, after starting to record the clearance time, the participants can tour and explore the three levels on their own, learn bronze knowledge, and then play the game. When the participants have questions, we provide timely answers. However, when the participants are tested on bronzeware knowledge in the game scene of each level, we do not provide any tips advice or answers regarding the relevant bronzeware knowledge. When a participant successfully passes the third level of the game, we stop the timer, and the participant has completed the experience of the bronze digital museum curation game.

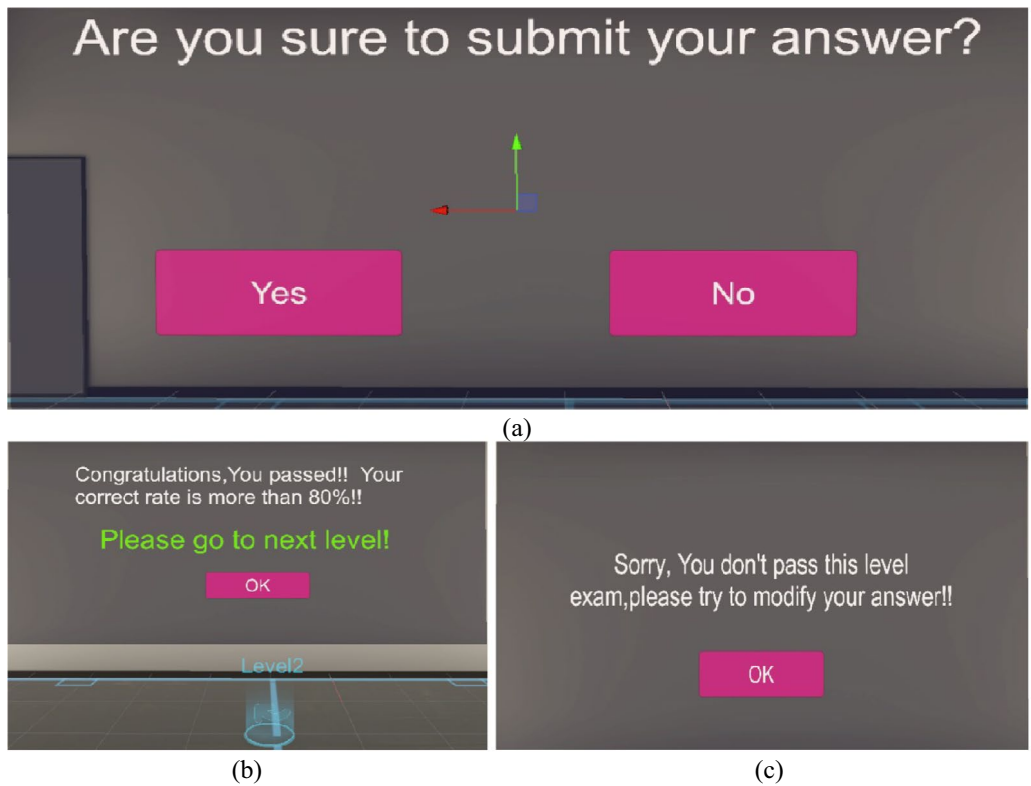

**Fig. 9** Answer submitting. **a** Correct rate check interface **b** Pass interface **c** Not pass interface

When the participants completed their experience, we used the SUS questionnaire [47, 48] to allow them to provide feedback based on a 5-point Likert scale [49] to rate our system. After all of the participants completed the questions and submitted them, we calculated each participant's score according to the scoring rules of the SUS questionnaire, and then convert the score into a hundred-point score. We received a total of 40 valid questionnaires and counted all of the SUS scores of the 40 participants.

**Table 2** SUS Questionnaire

| Questions |
| --- |
| I think that I would like to use this system frequently. |
| I found the system unnecessarily complex. |
| I thought the system was easy to use. |
| I think that I would need the support of a technical person to be able to use this system. |
| I found the various functions in this system were well integrated. |
| I thought there was too much inconsistency in this system. |
| I would imagine that most people would learn to use this system very quickly. |
| I found the system very cumbersome to use. |
| I felt very confident using the system. |
| I needed to learn a lot of things before I could get going with this system. |

The average SUS score was 77.3. The questions in the SUS questionnaire are shown in Table 2.

According to the research results of Sauro et al. [50] in 2011 on a database composed of SUS feedback data from more than 5,000 users in 500 studies, the SUS score can be converted into a percentile grade. The percentile grade measures the degree of availability of a system relative to other systems in the database. Our system's average SUS score converted to a percentile scale is approximately 81 points, which is a rating of B+ and means that our system's availability is higher than 81% of products or systems tested by SUS. In addition, according to Sauro et al.'s system usability measurement theory [51], the SUS questionnaire can also be divided into two parts for subscaling, that is, the part called "Learnability" is composed of questions 4 and 10 and the part called "Usability" is composed of the remaining 8 questions. These are used for a more specific and precise system assessment. By calculation, the average SUS score of the "Learnability" part was 59.5 points, which is approximately 30 points when converted to a percentile level, with a rating of C-. This shows that our system is easier to learn than 30% of the tested products or systems. The average score of the "Usability" part is 81.5 points, which is approximately 92 points when converted to a percentile level, and is rated A. This score indicates that the usability of our system is greater than 92% of the products or systems tested using the SUS.

In addition to using the percentile rank to explain the SUS score, we compared the SUS average score, acceptability score, adjective rating and the grade scale by Aroan Bangor et al. [52] in 2009, and the average SUS score of our system is within the acceptable interval, and the corresponding adjective rating is "Good".

The statistical results of each question in the SUS questionnaire are shown in Fig. 10. The results of Q1 show that the vast majority of participants want to use the system frequently. The results of Q3 indicate that the vast majority of participants think that the system is easy to use. The results of Q5 show that the vast majority of participants feel that the various functions of the system are well combined. The results of Q7 are that the vast majority of participants feel that most people can quickly learn to use this system. The results of Q9 indicate that the vast majority of participants are confident that they can use our system. The results of Q2 show that most participants feel that the operational design of this system is not complicated. The results of Q4 establish that approximately half of the participants needed the assistance of professionals. The Q6 results indicate that most participants feel that this system is consistent. The Q8 results show that most participants feel that this system is not cumbersome. The results of Q10 demonstrate that most people feel that they do not need to learn much before using the system.

To further verify the effectiveness of the VR-gamified virtual museum curation system(GVMCS) thst we built, we compared this system with its 2D desktop version of the virtual museum curation system (2DVMCS) which does not have the game scenes that reflect the advantages of our system in helping visitors better learn bronze knowledge and to deepen their ability to remember what they have learned. In this comparative experiment, we invited 40 participants and divided them into two groups, each with 20 people. The first group experienced the VR bronze gamified curation system, and the second group experienced its 2D desktop version of the curation system. The first group consisted of 13 masters degree students and 7 doctoral students, including 12 majoring in computer science, 5 majoring in chemistry, and 3 majoring in archaeology. The second group consisted of 14 masters degree students and 6 doctoral students, including 10 majoring in computer science, 7 majoring in archaeology, and 3 majoring in chemistry. The process of the participants in the first group was the same as that of the participants in the usability evaluation experiment, except they did not complete the SUS questionnaire. The experience of the participants in the second group was as follows: first, they are introduced how to move, adjust the perspective, view bronzes, etc. through the mouse and keyboard. The participants can then tour the three roaming scenes. When the participants indicate that they have finished the tour, the experience has been completed.

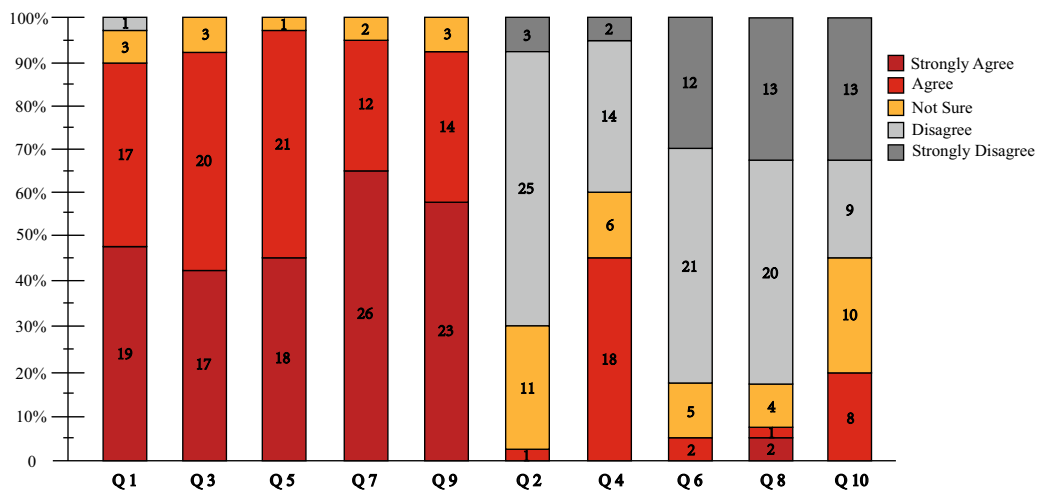

**Fig. 10** Results of each question in the SUS questionnaire

Each participant in the two groups needed to take a bronze knowledge test after experiencing the system, and the test questions for the two groups were the same. After the two groups of participants finished answering, SPSS software was used to test and analyze whether there was a significant difference in the test scores of the two groups of participants. Because the number of score data in each group was less than 50, the Shapiro-Wilk normality test was used to test the normality of each group's score data. After normality testing, we found that the test scores of the two groups did not conform to a normal distribution, non-parametric Mann–Whitney U test was used to perform a significant difference test. The test result was a p value less than 0.05, which showed that there was a significant difference in the test scores of the two groups. Through the two indicators of rank mean and rank sum, the scores of the participants using the VR GVMCS were higher than those of the participants using the 2DVMCS. The Shapiro-Wilk normality test result for the scores of the two groups is displayed in Table. 3. the result of the Mann–Whitney U test is shown in Table. 4 and Table. 5.

## Discussion and limitations

The results of the usability evaluation experiment and comparative experiment indicate that, first, our VR-gamified virtual museum curation system has strong usability, its operation is simple and easy to use, and the combination of functions in the system is natural and unobtrusive. Second, the system was well liked by the participants, and the participants were willing to use it regularly. It can enhance the visitor experience, and visitors are willing to immerse themselves in the tour and explore the bronze culture. Third, judging from the statistical results of the SUS subscale "Usability" part, compared with other systems, the usability of our system is excellent. Its usability is greater than that of systems tested with SUS. Finally, compared with the traditional 2D desktop version of the virtual museum curatorial system, the VR-gamified virtual museum curatorial system we designed with a level-breaking game link has significant advantages in helping visitors learn bronze cultural knowledge and deepen their ability to remember what they have learned.

Practice has shown that our VR game-based virtual museum curation system designed with the combination of virtual reality technology and "learn first, play later" interactive design can enable visitors to explore bronze models at any time and any place without the limitations of offline museums in terms of time, space, and distance and can stimulate their curiosity and desire to explore, give them an immersive experience and deepen their understanding of bronzes and their ability to remember related textual knowledge. In terms of cultural heritage, our system protects the precious Chinese bronze culture in a digitally and promotes the spread of traditional Chinese bronze culture, helping people around the world understand China's bronze culture. In addition, our research provides not only innovative design solutions for the popularization of bronze knowledge but also new methods and inspiration for peers to protect this cultural heritage.

Although the usability and effectiveness of this system are strong, there are still several shortcomings. First, from the perspective of "Learnability" in the SUS questionnaire's subscaling results, compared with other systems, our system still needs to be improved in terms of ease of learning, even though the results of Q7 of the SUS questionnaire show that the vast majority of participants(95%) felt that they could learn to use our system quickly. In the future, we will consider easier ways for visitors to understand and accept in interactive design, and strive to improve the average SUS score, so that the adjective evaluation of our system can be improved to "excellent". In addition, in terms of hardware, we used a handle controller, which cannot simulate the feeling of touching objects and finger operations in the real world, and a more realistic and accurate experience in such interaction for visitors could be provided. Finally, this system does not simulate the tour guide function in a real museum, thus, when visitors ask questions about bronze artifacts that require professional knowledge to answer, such as why this bronzeware is designed this way? How to read the name of a bronzeware? these questions cannot be answered by nonprofessionals, which can lead to a bad experience for visitors.

**Table 3** Results of the Shapiro-Wilk normality test

| Museum system | Statistic | df | .Sig |
| --- | --- | --- | --- |
| GVMCS | 0.518 | 20 | <0.001 |
| 2DVMCS | 0.895 | 20 | 0.033 |

**Table 4** Results of the Mann–Whitney U Test

| Museum system | N | Mean Rank | Sum of Ranks |
| --- | --- | --- | --- |
| GVMCS | 20 | 26.88 | 537.50 |
| 2DVMCS | 20 | 14.13 | 282.50 |

**Table 5** Results of the test statistics in Mann–Whitney U test

|  | Point |
| --- | --- |
| Mann-Whitney U | 72.500 |
| Wilcoxon W | 282.500 |
| Z | −3.500 |
| Asymp. Sig (2-tailed) | <0.001 |
| Exact Sig. | <0.001 |

**Conclusions and future work**

To stimulate visitors' interest and curiosity in exploring bronzeware and help them better learn and remember relevant knowledge about bronzeware, in this work, we digitize ancient China's precious bronze cultural heritage, design a digital museum environment with interactive behaviors, and create an interesting VR game. After the system was constructed, we invited students and teachers to experience it. Through feedback from the questionnaire, we find that our system is effective for allowing tourists to gain the knowledge of bronze culture and deepen their ability to remember what they have learned. We aim to help more tourists understand bronzeware through this VR game system while increasing the spread of bronzeware culture. In addition, we hope that our solutions for the protection and preservation of valuable cultural heritage such as digitized cultural heritage and the combination of VR games and learning can be provided to researchers in related fields.

With the continuous development of artificial intelligence and its popularity and application in society, it is becoming increasingly prevalent. In the future, we plan to add artificial intelligence elements to this VR game, such as combining 3D character models with large large models and retrieval-augment generation to turn the 3D character model into a dynamic tour guide non-player character(NPC), so that visitors can directly talk to the NPC through voice or text. This NPC will flexibly answer tourists' questions while enhancing the fun of the game and visitor's experience. For example, when a visitor enters the roaming scene, the NPC will follow the tourist as a tour guide. When a visitor is viewing the 3D bronze model and has questions while exploring, he can have a personalized conversation with the dynamic NPC guide; in this way, a personalized gaming experience can be provided for each visitor. We also plan to use network technology to enable multiple people to roam the bronze digital museum online in real-time, play VR games and use voice communication and other functions to create a VR-based metaverse system to achieve cooperation, sharing, and an exchange of experiences among tourists from different geographical locations and cultures. In terms of VR hardware, we plan to use haptic gloves to replace hand controllers in the future to achieve a more precise touch and interactive experience. In addition, in terms of the composition of participants who evaluate the effectiveness of the system, most of the participants in this work were college students, and the composition of the participants was relatively simple. We plan to add participants of different ages in future work to obtain a more general evaluation result.

We created a demo video to show our gamified bronze museum curation. The link is Gamified bronze museum curation video

**Abbreviations**

| | |
|---|---|
| 3D | Three-dimensional |
| BTFW | Break the fourth wall |
| AI | Artificial intelligence |
| HMD | Head mount display |
| VR | Virtual reality |
| UI | User interface |
| AR | Augmented reality |
| NPC | Non-player character |
| URP | Universal render pipeline |
| PBR | Physically based rendering |
| SUS | System usability scale |

**Acknowledgements**
We would like to thank Miao Tang and the staff at the Archaeology and Art Museum of Jilin University for their cooperation in helping us scan the bronzes.

**Author contributions**
LZK scanned the bronze, obtained the 3D bronze model data and processed the model. Then he built a VR digital museum, placed the game objects, designed the interactive behaviors and VR curation games,wrote this paper and prepared Figures 1–11 and Tables 1–2. ZQ scanned the bronze and provided proposals for curating games and bronzeware texts knowledge. XJY also proposed the macroscopical design of the project. YX and LCT proposed this project,provided guidance, and revised this paper.

**Funding**
This work is a phased achievement of the National Social Science Foundation project "Research on Chinese Pre Qin Language and Culture Based on Jinwen Data"(number:23VRC033). It is supported by the Youth Project of the Philosophy and Social Science Research Innovation Team of Jilin University "Inheritance of Chinese Character Culture and New Forms of Human Civilization"(number:2023QNTD02), and Jilin University (Grant No.419021422B08).

**Data availibility**
The game system's data are available and they can be requested from the corresponding author.

## Declarations

### Competing interests
The authors have no competing interests.

**Publisher's Note**